\documentclass[twocolumn,showpacs,floatfix,preprintnumbers,amsmath,amssymb,aps,pre]{revtex4-1}
\usepackage{amsmath}
\usepackage{amssymb}
\usepackage{graphicx} 
\usepackage{bm}
\usepackage{psfrag} 
\usepackage{color} 
\usepackage{graphicx}
\usepackage{pstricks}
\usepackage{pstricks-add} 

\newcommand{\lef}{\left(}						
\newcommand{\rig}{\right)} 					

\newcommand{\Dxi}[0]{\mathcal{D}_{\xi}}
\newcommand{\ns}[0]{\negthickspace}
\newcommand{\refeq}[1]{eq.\ (\ref{#1})}

\newcommand{\m}[1]{\ensuremath{\left< #1\right>}} 
\newcommand{\bs}[1]{\boldsymbol{#1}}
\newcommand{\fig}[1]{Fig. \ref{#1}}

\begin{document}

\title{Directed Transport of Confined Brownian Particles with Torque}

\author{Paul K. Radtke and Lutz  Schimansky-Geier}  
	
\affiliation{Department of Physics, Humboldt-Universit\"at zu Berlin, Newtonstr. 15, 12489 Berlin, Germany}
  

\begin{abstract}
  We investigate the influence of an additional torque on the motion of
  Brownian particles confined in a channel geometry with varying
  width. The particles are driven by random fluctuations
   modeled by an Ornstein-Uhlenbeck process (OUP) with given
    correlation time $\tau_{c}$. 
    The latter causes persistent motion and is implemented as (i) 
    thermal noise in equilibrium and as (ii)
    noisy propulsion in nonequilibrium.  
    In the nonthermal process a directed transport
    emerges, its  properties are studied in detail with respect to the 
    correlation time, the torque and the channel geometry. Eventually, 
    the transport mechanism is
    traced back to a persistent sliding of particles along the 
    even boundaries in contrast to scattered motion at uneven or rough ones. 
\end{abstract}

\maketitle

\textbf{Introduction.} Nowadays, ratchet models are employed to
  explain a wide range of transport phenomena, in biophysics as well
  as in artificial devices \cite{reimann2002brownian,hanggi2009artificial}. 
Generally, the term ratchet refers to nonequilibrium phenomena that
can arise provided one of the space-time symmetries that inhibits
directed motion is broken \cite{flach2000directed}. This can be caused
either explicitly by an asymmetric, periodic structure or by an
unbiased, but asymmetric drive \cite{goychuk2001minimal}. In our case,
the symmetry breaking is realized by a nonvanishing mean torque
together with the confining geometry.


The torque leads to a circular motion and an alteration of 
 the diffusive properties \cite{weber2011active}.  Such motion appears in many real
world phenomena. For example, it can be due to asymmetries in the
propulsion of agents themselves as occurring in the chemotaxis of
sperm cells
\cite{friedrich2008stochastic,friedrich2010high,vanteeffelen2008dynamics}
and nanorods \cite{sen2009chemo,ozin2010nanolocomotion}. Also, a
torque appears for Janus particles under laser irradiation
\cite{jiang2010active} or due to an external magnetic field that is
used to steer nanorods \cite{kline2005catalytic}.

\begin{figure}
\centering
\includegraphics[scale=1]{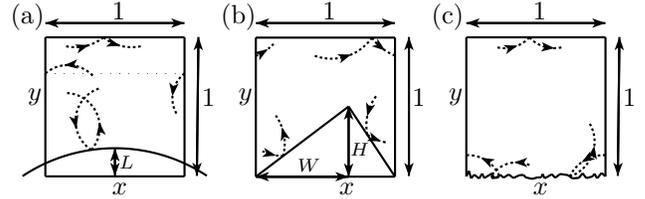} 
\vspace{-6pt}
\caption{Illustration of the channel geometry, a box of unit size
  whose left and right side are identified with each other, realizing
  periodic boundary conditions. At the top and bottom specular
  reflections take place. (a) Lower boundary formed by a disk
  of radius $R=1.2$ that protrudes by $L=0.2$ into the box. (b) Edged boundary 
parameterized by $H=0.5$ and $W=2/3$. (c) Flat
  lower boundary with equidistributed scattering angles due to a
  `rough' surface.}
\label{geometry}
\end{figure}

Furthermore, our particles are confined in an infinitely long channel
with a periodically varying width, see \fig{geometry}. One channel
boundary is introduced as a reflecting disk. Later on it will be interchanged
by a reflecting triangle, thereby adding another symmetry
breaking. Finally, we also consider a `rough' lower wall where elastic
scattering is modeled by equidistributed reflection angles, regardless
of the angle of incidence.

Even though related studies of transport mechanisms possess some of the
features, such as the channel layout in chaotic transport in
Hamiltonian systems \cite{schanz2005directed,acevedo2003directed}, entropic particle
transport \cite{martens2011entropic} or for confined biological agents
\cite{vanteeffelen2008dynamics}, its combination with a dissipative dynamics and a nonzero mean
torque to realize directed transport is a novel feature. Hereby, the actual driving of the particles is 
performed by the noise, which delivers a constant energy supply.

\begin{figure*}
\begin{center}
\includegraphics[scale=0.9]{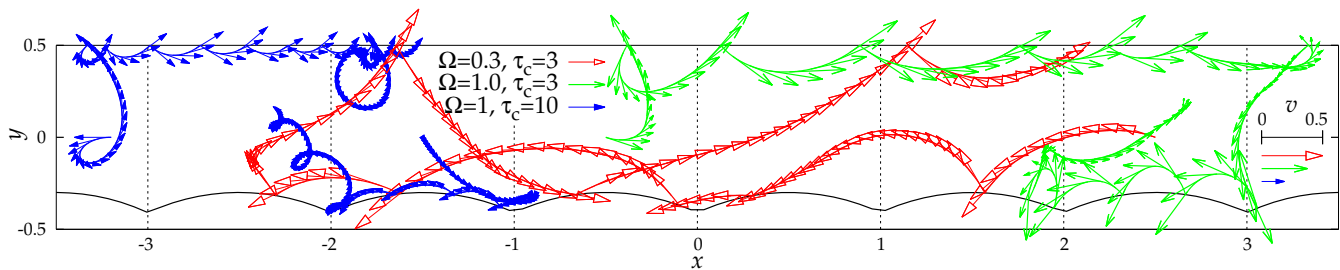} 
\vspace{-16pt}
\end{center}
\caption[Trajectories with colored noise in a confined
geometry]{(Color online) Trajectories of confined (cf. \fig{geometry}
  (a)) particles exposed to nonthermal colored noise with several
  torques and correlation
  times 
  at constant $\gamma_0=0.2$, $\mathcal{D}_{\xi}=0.04$. The depicted
  time length is $t_{l}=33$, the velocity is indicated by the arrows,
  between two consecutive arrows $\Delta t=1/4$ has passed.}
\label{coloredtrajectories}
\vspace{-6pt}
\end{figure*}

\textbf{Thermal Colored Noise.} In our setting the particle has unit
mass $m=1$, position $\bs{r}(t)=(x(t),y(t))$ and velocity
$\dot{\bs{r}}(t)=\bs{v}(t)=(v_{x}(t),v_{y}(t))$ at time $t$, that are
driven by correlated noise
$\bs{\epsilon}(t)=(\epsilon_{x}(t),\epsilon_{y}(t))$. Such noise
  is called colored, its memory causes a persistent
  motion. We choose an OUP to drive our particles, which is a common
  implementation for it represents the most natural continuous valued
  colored noise \cite{hanggi1995colored}. 
In equilibrium a  fluctuation dissipation relation (FDR) relates the 
autocorrelation function of the noise and the friction function
\begin{equation}
\m{\epsilon_{i}(s)\epsilon_{j}(s+t)}= \frac{\Dxi}{\tau_{c}}e^{-|t|/\tau_{c}}\delta_{ij}= kT \gamma(|t|)\delta_{ij},
\label{corr-epsilon}
\end{equation}
with $i,j \in\{x,y\}$. Here the Einstein relation $\Dxi=\gamma_0 k T$ ($\gamma_0$ --
friction constant, $k$ -- Boltzmann constant, $T$ -- temperature) is
used to relate the noise intensity $\Dxi$ to the heat bath from which the
noise derives. Assuming that a FDR holds, we are forced to relinquish
the Markov property by introducing a dissipative memory kernel
$\gamma(|t-s|)$ to govern the friction.  We then speak of a
generalized Langevin equation \cite{mori1965continued,paraan2008brownian}
 for the velocity,
\begin{equation}
\dot{\bs{v}}(t) = -\ns\int_{0}^t\ns\gamma(t-s)\bs{v}(s)\text{d}s +\bs{\epsilon}(t) +  \bar{\bs{\Omega}}\bs{v}(t).
\label{col-int}
\end{equation}
The mean torque is realized by multiplication of $\bs{v}$ with the
rotation matrix $\bar{\bs{\Omega}}$, resulting in the force
$\bar{\bs{\Omega}}\bs{v}=(\Omega v_{y}, -\Omega v_{x})$.

Without confinement, the stationary Fokker-Planck equation
corresponding to this dynamics is easily solved. We then find that the
velocity is Maxwell-Boltzmann distributed. Numerically, we can confirm
that this holds true in the channel geometry as well, both globally
for the entire channel and in the vicinity of the boundaries. Hence we
have a symmetrical velocity distribution and no directed transport
occurs.

The  particles perform an undirected diffusive motion.  As an estimate one can take
  the effective diffusion coefficient of this dynamics calculated for
unbounded space, which reads \cite{paraan2008brownian}
\begin{align}
D_\text{eff} &=  \Dxi/\lef\gamma_0^{2}+\Omega^{2}\rig.
\label{deff-coleq}
\end{align}
Apparently, it does not depend on the correlation time and coincides
with the expression for particles driven by white noise.

\textbf{Nonthermal Colored Noise.}
\label{Nonequilibrium Colored Noise}
In the second case we consider noisy propulsion in nonequilibrium, hence no FDR holds
\cite{hwalisz1989colored}. Again equipped with an OUP-driving, our
system is governed by the set of Langevin equations
\begin{align}
\begin{aligned}
\dot{\boldsymbol{r}}(t)		=	& \boldsymbol{v}(t), \qquad 
\dot{\boldsymbol{v}}(t)		= 	-\gamma_0 \boldsymbol{v}(t) + \bar{\boldsymbol{\Omega}}\boldsymbol{v}(t) + \boldsymbol{\epsilon}(t), \label{eom-col2}\\
& \dot{\boldsymbol{\epsilon}}(t) = -\frac{1}{\tau_{c}}\boldsymbol{\epsilon}(t) +\frac{\sqrt{2\mathcal{D}_{\xi}}}{\tau_{c}}  \boldsymbol{\xi}(t).
\end{aligned}
\end{align}
Here $\boldsymbol{\xi}(t)$ denotes white Gaussian noise of zero mean
with uncorrelated components, i.e.  $\m{\xi_{i}(t)}=0$ and
$\m{\xi_{i}(s)\xi_{j}(s+t)}=\delta(t)\delta_{ij}$. 

In this way, we have effectively realized the motion of an
  active particle, it underlies a permanent correlated
  supply of energy from the OUP-noise which it dissipates during the
  persistent and curved motion due to Stokes friction. The persistence
  of this motion expresses nonequilibrium.  With $\tau_{c}\rightarrow
  0$ white noise follows realizing again an equilibrium situation. 

By taking the Fourier transforms of \refeq{eom-col2}, the spectrum of
the velocity can be traced back to that of the noise
\begin{equation}
S_{v_{i}v_{j}}(\omega) =
\frac{(\omega^2+\gamma_0^2+\Omega^2)}{(\gamma_0^2+\Omega^2-\omega^2)^2+4\omega^2\gamma_0^2}S_{\epsilon_{i}\epsilon_{j}}({\omega}).
\label{velautocorr-general}
\end{equation}
With the Wiener-Khintchine
theorem 
and the Lorentzian noise spectrum of the OUP
 $S_{\epsilon_{i}\epsilon_{j}}(\omega) 
= 2 \mathcal{D}_{\xi}\lef 1+\tau_{c}^{2} \omega^2\rig^{-1} \delta_{ij}$,
which follows from the left-hand side of \refeq{corr-epsilon},
we obtain a rather lengthy expression for 
$\m{v(s)v(s+t)}$.
For $t=0$ it simplifies to
\begin{equation}
\m{v_{i}v_{j}} =  \frac{\mathcal{D}_{\xi}}{\gamma_0}\frac{\gamma_0\tau_{c}+1}{(\gamma_0\tau_{c}+1)^2+\Omega^2\tau_{c}^{2}}\delta_{ij}.
\label{eq:vsquare_t=0}
\end{equation} 
Obviously, the torque, the friction, and the correlation time dampen
the particles' mean squared velocity. For a vanishing mean torque the
result without external force field \cite{hwalisz1989colored} is
reproduced.

We notice that the unconfined effective diffusion coefficient can be
calculated from the Fourier transform of \refeq{velautocorr-general}
by integration over the time,
which leads to the same result as for thermal colored noise
(cf. \refeq{deff-coleq}).

\textbf{Directed Transport in Confined Geometries.}
\fig{coloredtrajectories} depicts some trajectories in the circular
channel for the nonthermal dynamics given by \refeq{eom-col2}. Rising
correlation times lead to a reduction of the total spatial
displacements as do increasing mean torques, although less
severely. With both influences acting on the particles together,
another effect can be observed: The particles stay in the vicinity of
the reflecting boundaries for a longer time and perform a curly
hopping motion that changes to a narrow creeping for large $\tau_{c}$
\cite{sliding}. Also, we notice that while the red trajectory (empty arrowheads) is
largely unbiased, with bigger torques the particles travel a longer
distance along the flat wall than along the curved wall.

\begin{figure}
\centering
\includegraphics[scale=0.9]{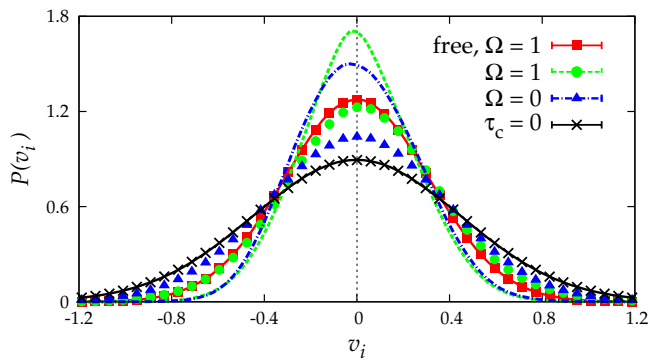}
\vspace{-6pt}
\caption[Velocity distribution for particles driven by colored
noise]{(Color online) Velocity distribution ($v_{x}$ -- symbols,
  $v_{y}$ -- lines) for particles driven by nonthermal colored noise
  with $\tau_{c}=1$, $\gamma_0=0.2$, $\mathcal{D}_{\xi}=0.04$.
  Except for the red distribution (squares), particles are
  confined as in \fig{geometry} (a). For $\tau_{c}=0$, the same Gaussian
  distribution emerges regardless of $\Omega$.} 
 \label{dist_color}
\vspace{-6pt}
\end{figure}

The velocity distributions for the same situation are depicted in
\fig{dist_color} \cite{mannella2000gentle}. While the velocity is
Gaussian distributed for $\tau_{c}=0$, it remains symmetrical
on an unbounded plane or in the confined geometry without constant
torque. In contrast, if all these influences act on the particles
together, i.e. boundaries, finite correlation time and constant torque
field, the velocity distribution loses its symmetry (green
circles). For $\Omega=1$, the $v_{x}$-distribution's left tail is
lowered, while its right tail is raised. Thus, we have a net particle
flux. Unlike for particles in equilibrium, the velocity distributions
are narrowing with growing $|\Omega|$.  Furthermore, we notice that
the confinement particularly leads to a narrowing in $P(v_{y})$, which
now markedly differs from $P(v_{x})$.

Let us now address the stationary particle fluxes $J$ that pass in the
$x$-direction through our channel. Herein, $J$ is normalized with
respect to the number of particles.

\begin{figure}
\centering
\begin{centering}
\includegraphics[scale=0.9]{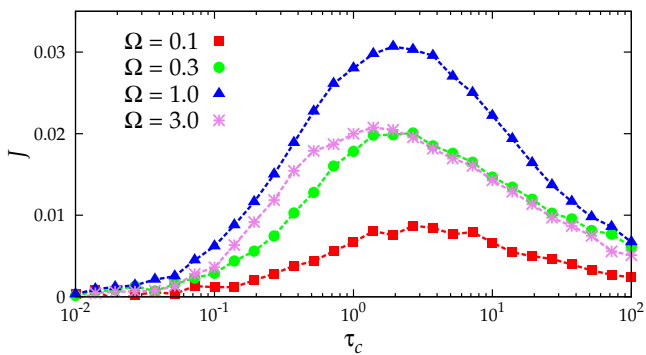}
\end{centering}
\vspace{-6pt}
\caption[Net transport for particles driven by colored noise as a function of the correlation time]{(Color online) Net transport for particles driven by nonthermal colored noise for several mean torques at $\gamma_0=0.2$, $\Dxi=0.04$. The channel geometry is specified in \fig{geometry} (a).}
\label{flow_color_tau}
\vspace{-6pt}
\end{figure}

The numerical results for the net transport are shown in
\fig{flow_color_tau} as a function of the correlation time for several
values of the mean torque. The flux $J$ has a maximum at approximately
$\tau_{c}\approx 2-3$ for all depicted $\Omega$
values. 
For vanishing correlation times, $J$ converges to zero, which is
consistent in view of the white noise case that follows in this
limit. There, the equilibrium condition forbids directed transport as
the noise strength in \refeq{eom-col2} vanishes. For large correlation
times the net transport also converges to zero.

\begin{figure}
\begin{centering}
\includegraphics[scale=0.9]{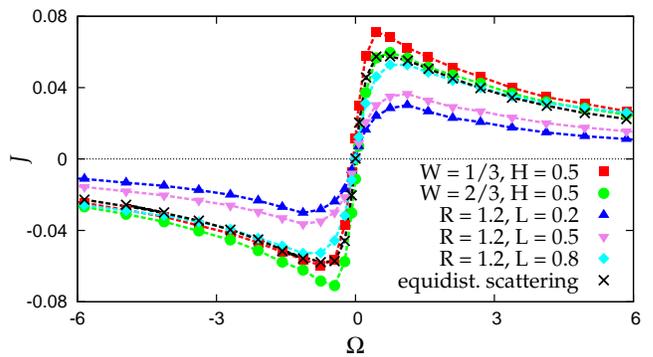}
\end{centering}
\vspace{-4pt}
\caption[Net transport for particles driven by colored noise as a function of the mean torque for various confining geometries]{(Color online) Net transport for various confining geometries (cf. \fig{geometry}). All curves are at $\gamma_0=0.2$, $\Dxi=0.04$ and  $\tau_{c}=3$. }
\label{flow_color}
\vspace{-6pt}
\end{figure}

The effect of the mean torque is examined in further detail for
several geometries in \fig{flow_color}. Evidently, $J$ vanishes for
$\Omega=0$ and also for strong torques regardless of the mean turning
direction. The direction of the net transport changes with the
orientation of $\Omega$.

Given the emergence of a directed transport, we can justify its
qualitative form, i.e.\ the asymptotics in $\tau_{c}$ and $\Omega$ as
well as the occurrence of a maximum, with regard to the behavior of
the speed. As implied by \refeq{eq:vsquare_t=0} it vanishes both for
large mean torques and correlation times. For transport to take place
however, we need a torque to break the symmetry, and a nonzero
correlation time for nonequilibrium. Hence we have no transport in
either of the cases $\Omega\rightarrow 0$ or  $\Omega\rightarrow\infty$ and $
\tau_{c}\rightarrow 0$ or $\tau_{c}\rightarrow\infty$, leaving only a
maximum value in between, which has to change its direction with the sign
if the sign of the torque switches.

We now turn to the effects of the differing geometries in more
depth. As we see in \fig{flow_color}, the antisymmetrical behavior
in $\Omega$ is conserved only as long as the lower boundary keeps as
left-right symmetry (cf. \fig{geometry} (a,c)). Otherwise, as for the
edged boundaries (cf. \fig{geometry} (b)), the heights of the positive
and negative peaks differ, whereas the direction of the current
coincides for $W=1/3$ and $W=2/3$ (i.e. is independent of the
orientation of the sawteeth). 
Also, the position of the maximum moves to slightly smaller $\Omega$
values for the edged surface compared to the disk.

If we narrow the channel by increasing the origin of the reflecting
disk, the resulting transport curves also rise. However, their overall
behavior remains; in particular the position of the maximum is barely
altered and remains at $\Omega^{\text{max}}\approx 1-1.5$. From
this we conclude that the position of the $\Omega^{\text{max}}$
cannot be traced to some resonant-like effect between the channel
width and the cyclotron radius $R_{c}=v/\Omega$. Rather, the position
of $\Omega^{\text{max}}$, as well as $\tau_{c}^{\text{max}}$, is a
tradeoff between the necessary symmetry breaking and the reduction of
the particles' speed.

Due to the unit scaling of the velocity, our net flux is identical to
the mean velocity in the $x$-direction, $J=\langle v_{x} \rangle$. Other
correlation ratchets \cite{bartussek1996precise,lindner1999inertia}
display values for $\langle v_{x} \rangle$ in the same order of
magnitude.

\textbf{Elucidation of the Transport Mechanism.}  There are three
possible types of trajectories in our channel: particles in the middle
perform a more or less circular motion without net bias, whose
regularity is enhanced by $\Omega$ and decreased by $\Dxi$ and
$\gamma_0$. Particles near the top eventually collide with the wall
and are reflected. A counter clockwise mean turning rate then results
in a forward motion and multiple further reflections.  If we consider
a particle that is driven to the wall by a noise pointing into that
particular direction, the noise works largely as a drag force after
reflection. It is still oriented in the same direction, whereas the
particle has changed its own. As a result, the velocity normal to the
boundary decreases (cf.\ \fig{dist_color}). Once the noise orientation
has changed (after about $\Delta t\approx\tau_{c}$), the particle is
driven away from the boundary, although this may be counteracted by
the particle's inertia together with the torque. Such situations are
shown in \fig{coloredtrajectories}. Clearly visible are the repeated
reflections with ever decreasing radius and speed at the top.

Without additional obstacle, particles near the bottom perform an
antagonistic motion that cancels out the forward flux. The obstacle on
the other hand modulates the reflection angles depending on the point
of incidence. As seen in \fig{coloredtrajectories}, the distance the
particles have to travel until the next reflection varies depending on
whether the surface tangential is increasing or decreasing.  This
constitutes a scattering effect that helps particles to escape from
the vicinity of the
boundary. 

Hence the backward flux is hindered and a positive net transport
emerges. In this picture, we can also explain the augmentation of $J$
with the narrowing of the channel (cf.\ \fig{flow_color}), for the
particles cannot move as far without interaction with the
boundaries. Thusly, the proportion of intermediate unbiased motion is
reduced and the impact of the symmetry breaking is enhanced.

We confirm this rationale by considering a box with a flat but rough
lower boundary, modeled by random equidistributed reflection angles
(cf. \fig{geometry} (c) and \fig{flow_color}, black symbols).
Obviously, this leads to a similar behavior of $J$.

Consequently, the flux is dominated by the sliding motion along the flat
boundary and affected by the persistence, the torque and the noise
driven velocity. In contrast, the geometry of the confinement 
and the shape of the lower boundary appear to be less important and influence the total 
  value of $J$ rather than its form. 
  To evaluate qualitatively the dynamical dependence of the current, we expect
  that $J=J(\tau_c,\Omega,|{\bf v}|)$. It has to vanish if any of these
  three parameters vanishes, which favors a product of them. We can assume proportionality
  to the velocity, which we approximate by $\sqrt{\m{\bs{v}\bs{v}}}$. Then, the dependence 
  from the torque is met by an additional factor $\sqrt{\Omega}$. The flux $J$ rises linearly 
  for small $\tau_c$, in comparison to the numerical results this must be supplemented by an additional damping 
  for large ones. Hence we propose $J\propto \text{sgn}(\Omega)\sqrt{ |\Omega|
  \m{\bs{v}\bs{v}}}\tau_{c}\exp(-\tau_{c}/4)$, which mimics the numerical results 
  quite well. It yields a optimal torque at $|\Omega^\text{max}| =
  \tau_{c}^{-1}+\gamma$, where the transport has a maximum of $J^{\text{max}} \propto \tau_c
  \exp({-\tau_{c}/4})$. This behavior can also be seen in 
  \fig{flow_color_tau}, 
   where
   $\Omega^{\text{max}} $ wanders to slightly smaller values
  with $\tau_{c}$, and in the height of the corresponding peaks. It can
  be interpreted as a time scale merging between the period due to the
  torque, which is proportional to $1/\Omega$, and the two relaxation times
  $\tau_c$ and $\tau_\text{inertia}=1/\gamma$ that characterize the
  persistence of the motion.


In conclusion, we have constructed a setup in which unbiased
correlated noise is rectified in nonequilibrium and leads to the
emergence of directed transport. To that end, we needed a nonzero mean
torque and an asymmetric confining channel. We have given a
qualitative explanation of the mechanism that leads to this symmetry
breaking, namely the particle hopping along the boundaries disrupted
by the scattering effect of the reflecting disk or triangle. Despite the
general settings of the propulsion, particles and geometry, this
investigation can help in the qualitative understanding and might be
the starting point for more precise application oriented modeling of noisy active
particles.

\textbf{Acknowledgments.} This paper was developed within the
  scope of the IRTG 1740 funded by the DFG.  We acknowledge fruitful
  discussions with C. van den Broeck, T. Dittrich and B.
  Sonnenschein.


\begin{thebibliography}{22}
\bibitem [{\citenamefont {Mannella}(2000)}]{mannella2000gentle}%
  \BibitemOpen
  Errorbars are of symbol size and hence omitted. \BibitemShut {Stop}
\end{thebibliography}

%

\end{document}